\documentclass[3p,twocolumn,sort&compress,review]{elsarticle}

\usepackage{amsmath}
\usepackage{amssymb}
\usepackage{gensymb}
\usepackage{xfrac}

\usepackage{graphicx}
\usepackage{subfig}
\usepackage{verbatim}

\usepackage{url}

\usepackage[colorlinks=true, allcolors=blue]{hyperref}

\usepackage{cleveref}

\newcommand{\vud}{$V_{ud}$}

\providecommand{\urlprefix}{URL: }

\begin{document}

\begin{frontmatter}

\title{Off-line Commissioning of the St.~Benedict Radio Frequency Quadrupole Ion Guide}

\author[ND]{R.~Zite\corref{cor1}}\ead{rzite@nd.edu}
\author[ND]{M.~Brodeur}
\author[ND]{O.~Bruce}
\author[ND]{D.~Gan}
\author[ND]{P.D.~O'Malley}
\author[ND]{W. S.~Porter}
\author[ND]{F.~Rivero}

\address[ND]{Department of Physics and Astronomy, University of Notre Dame, Notre Dame, IN, USA}

\cortext[cor1]{Corresponding author}

\begin{abstract}
The Superallowed Transition Beta-Neutrino Decay Ion Coincidence Trap (St.~Benedict) is currently under construction at the Nuclear Science Laboratory (NSL) of the University of Notre Dame. It aims to measure the beta-neutrino angular correlation parameter for superallowed mixed mirror beta decay transitions. Measurements of this kind offer unique insight into the electroweak part of the Standard Model through tests of unitarity of the Cabibbo-Kobayashi-Maskawa (CKM) matrix. St.~Benedict is comprised of several beam-manipulating components including a radio frequency quadrupole (RFQ) ion guide. This ion guide features an off-line source at 90\degree\ to the beam path for testing and calibration of downstream components once St.~Benedict is online. Off-line commissioning of the ion guide demonstrated a transport efficiency greater than 95\% for ions coming from the upstream RF carpet chamber. When taking ions from the 90\degree\ off-line source a lower efficiency of 60\% was obtained.
\end{abstract}

\begin{keyword}
Radio-Frequency Quadrupole, Radioactive Ion Beam Manipulation, Paul Trap
\end{keyword}

\end{frontmatter}


\section{Introduction}


To date, the Standard Model (SM) is the most complete theoretical description of fundamental particles and their interactions \cite{PDG}. However, it fails to explain the matter/antimatter asymmetry, dark matter, and does not include gravity \cite{PhysRevD.86.010001}. As such, there are many experimental efforts dedicated to searching for physics beyond the Standard Model (BSM). One such probe is the unitarity test of the Cabibbo-Kobayashi-Maskawa (CKM) quark mixing matrix \cite{Severijns_2013}. 

Recent improvements to a theoretical correction term  that enters into the determination of the largest element of the CKM matrix, \vud, have resulted in a $3\sigma$ tension with unitarity \cite{3sigma}. This has prompted the need to extract \vud\ from alternative superallowed beta decay transitions, including those between nuclear mirrors, which require the determination of the Fermi to Gamow-Teller mixing ratio, $\rho$. To that end, the Superallowed Transition BEta-NEutrino Decay Ion Coincidence Trap (St.~Benedict) \cite{Brodeur2016-StBenedict, OMalley2020-StBenedict} is currently being commissioned downstream of the \textit{TwinSol} facility at the Nuclear Science Laboratory (NSL) of the University of Notre Dame. It aims to extract $\rho$ via a measurement of the time-of-flight spectra of the recoiling daughter nuclei of several mirror isotopes ranging from $^{11}$C to $^{41}$Sc held in a Paul trap \cite{Porter2023-StBenedict}. 

Measurements of this kind at an in-flight facility such as \textit{TwinSol} requires the stopping of fast radioactive ion beams (RIBs) and the manipulation of the thermalized ions using several devices. One of these, the radio-frequency quadrupole (RFQ) ion guide, is commonly used for beam transport at low energies by employing fast oscillating electric fields to radially confine the beam in intermediate pressure regions ($\approx 10^{-3}$ Torr) \cite{SAVARD2020258, BLOCK20084521}.

This paper presents the off-line commissioning of the St.~Benedict RFQ ion guide in its final position in a differentially pumped chamber. The ion guide has been commissioned with an off-line ion source located upstream along its axis. Furthermore, an off-line source is located at 90\degree\ to the beam path intended for the testing and calibration of downstream components once St.~Benedict is online. The commissioning of the ion guide focused primarily on studying the effect of various static potentials, RF amplitude and pressure on transport efficiency to find possible limiting factors and mitigate them. This was done for both of the modes of transport described above.

\begin{figure*}
    \centering
    \includegraphics[width=\linewidth]{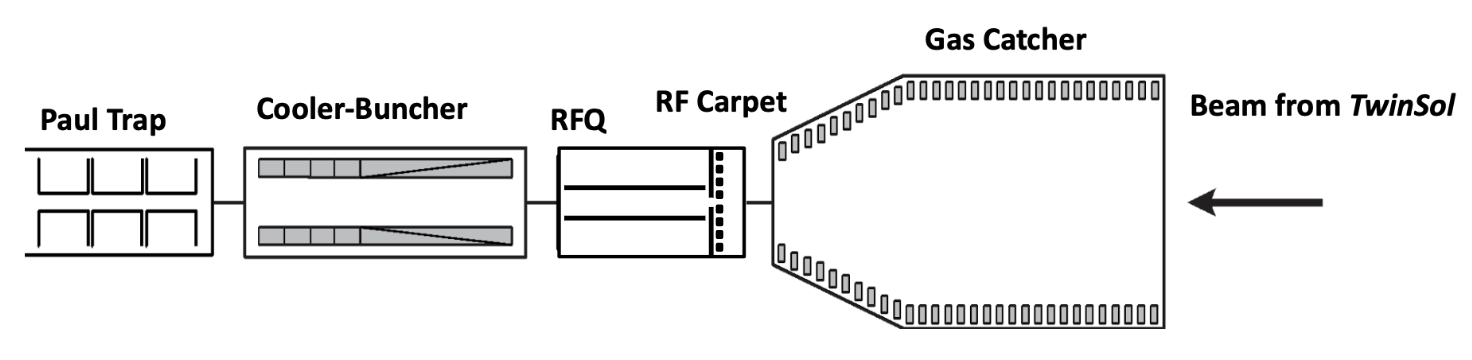}
    \caption{A diagram displaying each of the four main components of St.~Benedict}
    \label{fig:St.Ben}
\end{figure*}

\section{St.~Benedict}

St.~Benedict takes RIBs produced at the \textit{TwinSol} facility \cite{BECCHETTI2003377, OMALLEY2016417} and prepares them for injection into a measurement Paul trap. The primary objective of the various components upstream of the Paul trap consists in efficiently converting the fast continuous beam from \textit{TwinSol} into low-energy ion bunches. St.~Benedict, shown in figure~\ref{fig:St.Ben}, consists of four main elements. The first is a large volume gas catcher \cite{BRODEUR202379} designed to stop fast beam coming from \textit{TwinSol}. The ions then enter a differential pumping system \cite{Davis2022-Flow} consisting of two chambers housing a radio frequency (RF) carpet \cite{Davis2022-Static} and an RFQ ion guide for beam transport. The third element is an RFQ cooler-buncher \cite{CoolerBuncher} which creates cool ion bunches for injection into the last element, a measurement Paul trap \cite{Porter2023-StBenedict}. 

The St.~Benedict gas catcher typically requires, besides solid material, at least $30$ Torr of helium to efficiently thermalize the $10-40$ MeV beams produced by \textit{TwinSol}. However, to efficiently transport and inject the ion beam in the RFQ cooler-buncher, a pressure near $10^{-5}$ Torr needs to be achieved. To get to this pressure, differential pumping is required and accomplished using a chamber (shown in figure \ref{fig:setup}) divided in three sections. The first section operates at a pressure on the order of a few Torr and ions are transported through this section using an RF carpet. A complete description of the RF carpet, including its off-line commissioning, can be found in \cite{Davis2022-Flow}. The next section maintains a pressure near $10^{-2}$ Torr and contains the RFQ ion guide, the off-line commissioning of which will be the main focus of this paper. Finally the last section contains two cylindrical electrodes to focus the ions at the entrance of the cooler-buncher. Pressure in this region, which will ultimately share the same volume as the injection optics of the cooler-buncher \cite{CoolerBuncher}, is around $10^{-5}$ Torr. The main application requirement of the RFQ ion guide is efficient transport of ions from the RF carpet chamber to the cooler-buncher. As such, this manuscript reports transport studies aimed at maximizing the transport efficiency of the device. Potassium ions, which are at a comparable mass to isotopes we plan to transport once St.~Benedict is online, were used for the tests as well as the previously published commissioning of the RF carpet \cite{Davis2022-Flow,Davis2022-Static} and RFQ cooler-buncher \cite{CoolerBuncher}

\section{The St.~Benedict Radio Frequency Quadrupole Ion Guide}

\begin{figure}
    \centering
    \subfloat[]{\includegraphics[width = 0.7\linewidth]{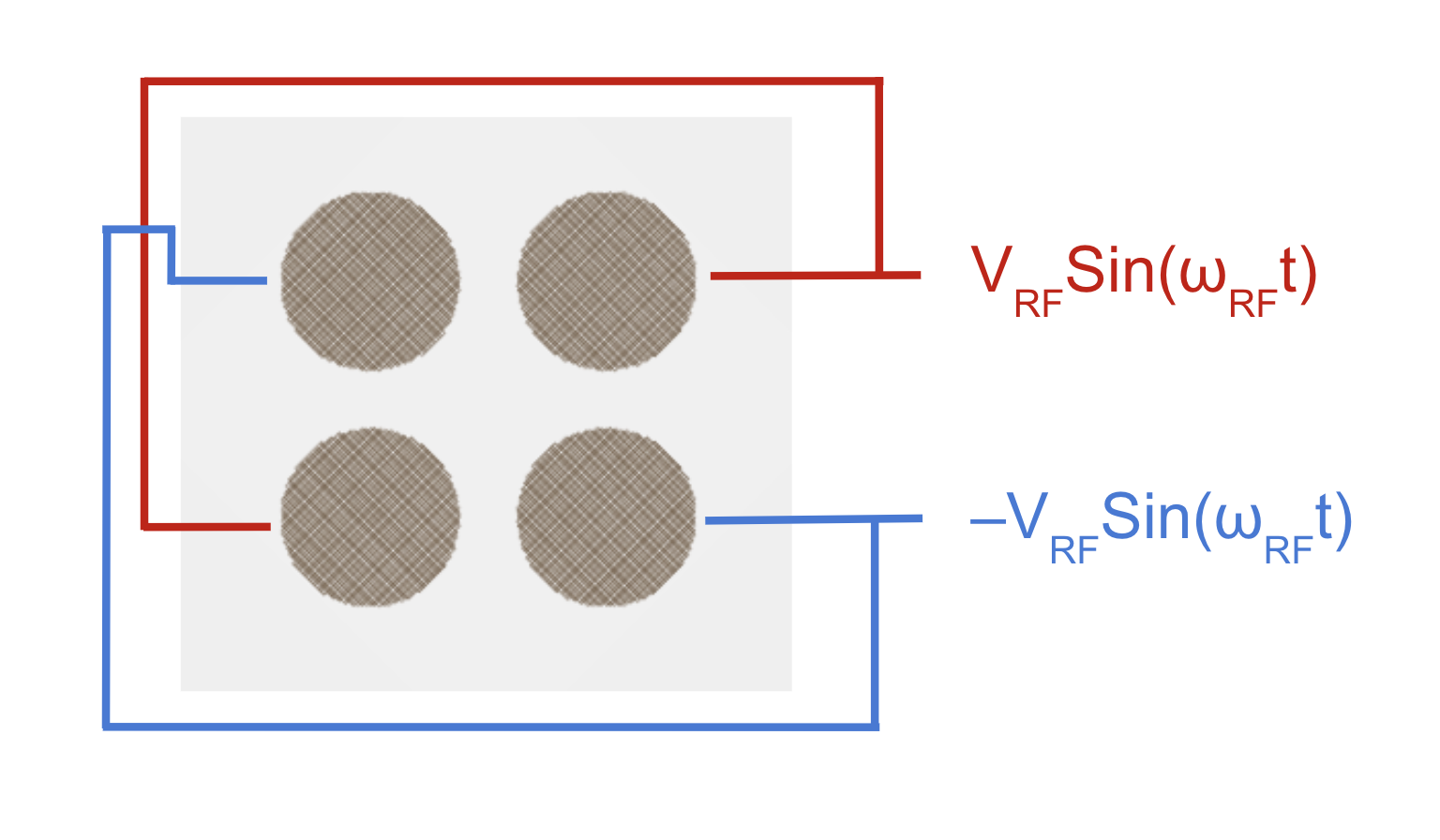}\label{fig:RFQFront}}\\
    \subfloat[]{\includegraphics[width = \linewidth]{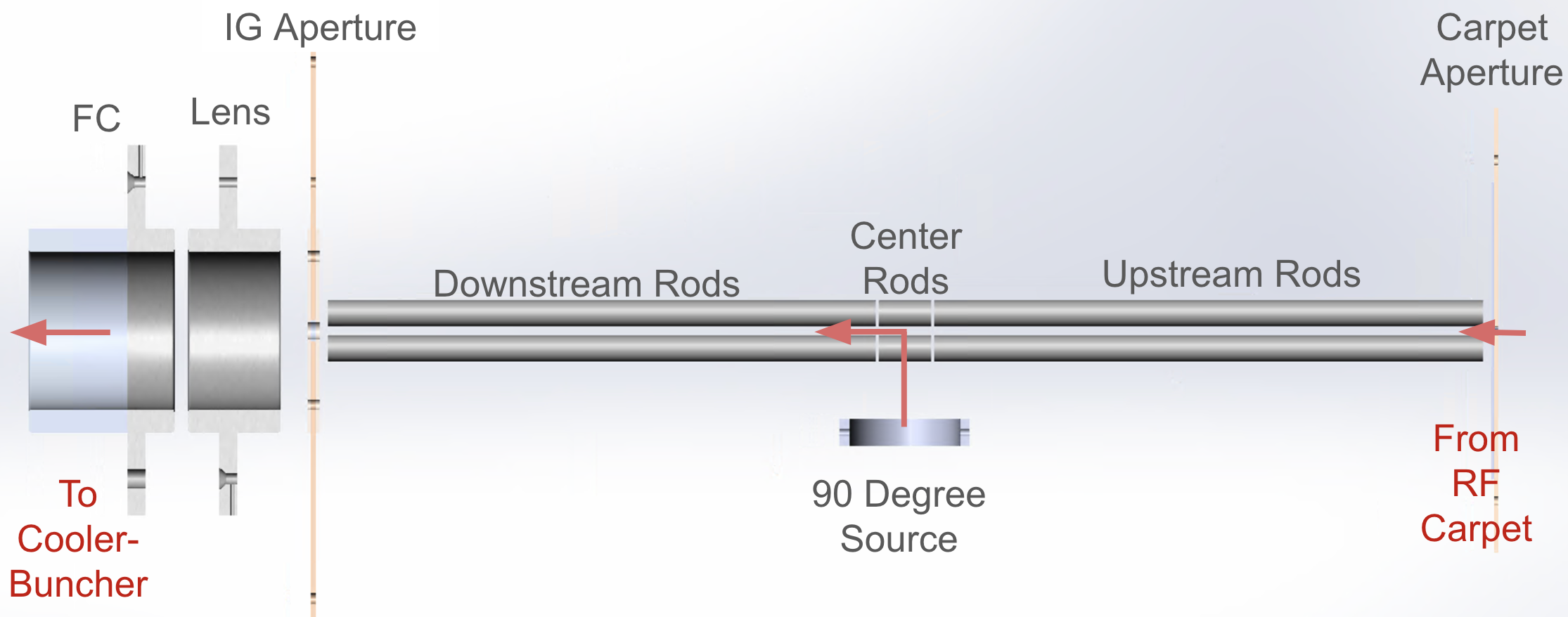}\label{fig:RFQLength}}
    \caption{A diagram of the St.~Benedict radio frequency quadrupole ion guide off-line commissioning setup.
    (a) Head on view displaying the quadrupole structure of the rods.
    (b) Side view of the RFQ ion guide displaying the segmentation of the rods, the 90\degree\ ion source, both apertures as well as the lens and Faraday cup electrodes. The supporting frame has been removed for clarity.}
\end{figure}

The ion guide consists of three segmented regions of a quadrupole rod structure (see figure~\ref{fig:RFQLength}) where the segment closest to the RF carpet will be referred to as the upstream rods, the small central section will be the center rods and the last section will be the downstream rods. Ions are radially confined in this electrode structure by applying a sinusoidal radio frequency signal with a 180\degree\ phase shift between adjacent electrodes as can be seen in figure~\ref{fig:RFQFront}. These ions are then transported by their original kinetic energy and the potential differences across the three sections. 

Space constraints in the Far West Target Room (FWTR) of the NSL where St.~Benedict will be located, proved the need for a unique solution for off-line testing and calibration of St.~Benedict components. As such, attached to the outer frame of the ion guide, at 90\degree\ to the beam path, is a thermionic emission potassium off-line source as described in section \ref{set up}. Circuitry was designed as described in section \ref{circuit} such that ions coming from this 90\degree\ source are able to enter into the ion guide before taking a 90\degree\ turn and exiting through the downstream aperture.

\subsection{Electrode Geometry}\label{electrodeGeometry}

Each of the three segments of the ion guide contains four stainless steel cylindrical rods with an $R = 3.21$ mm radius and an inscribed radius of $r_o = 3$ mm. These dimensions were chosen to closely match the ideal ratio $R/r_0 = 1.13$ \cite{Douglas2002} that minimizes non-harmonic components in the radial potential while using material of standard imperial sizes. There is a $1$ mm gap between each segment of rods. The outer two segments are longer, with the upstream rods measuring $187.8$ mm and the downstream rods measuring $187.4$ mm. The length of the center section was based on ion optical simulations aimed at optimizing beam transport from the 90\degree\ source and, as such, is much smaller at $18$ mm. 

\begin{figure}
    \centering
    \rotatebox{0}{\includegraphics[width=0.6\linewidth]{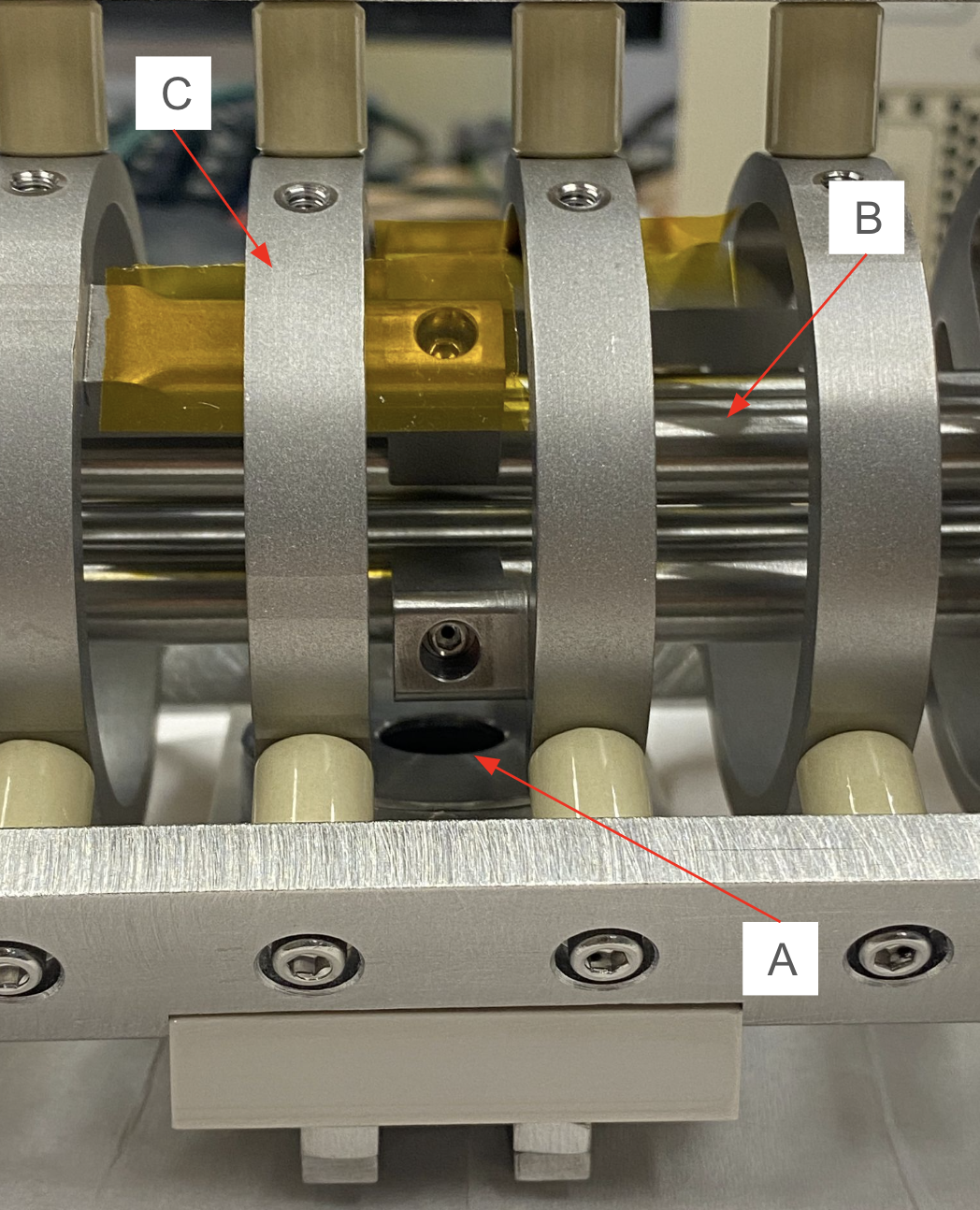}}
    \caption{(A) The 90\degree\ source as shown in reference to (B) the ion guide rods and (C) the supporting framework.}
    \label{fig:ion source}
\end{figure}

Attached to the frame, which holds the rods in place, and just below the center sections, is a holder for an off-line ion source. Simulations of the source holder at both $10$ and $20$ mm below the center rods both showed efficient transport through the downstream rods. Hence, we chose to place the ion source at a distance of 20 mm as it facilitated the mechanical design of the structure. This design places the source directly in the middle of the center section, approximately $20$ mm below the central axis of the ion guide as can be seen in figure~\ref{fig:ion source}.

\subsection{RF Circuitry}\label{circuit}

The RF circuitry was designed for operation in two different running modes: 0\degree\ and 90\degree. The 0\degree\ mode maintains the RF signal on all four rods of the center section for continuous radial confinement of ions coming from the RF carpet. The 90\degree\ mode allows for ions to enter the ion guide from the 90\degree\ source by removing the RF signal from the lower two rods in the center section of the ion guide while maintaining the RF signal on the upper two. In this configuration, a DC potential is still applied to each rod in the center section. This effectively bends the ions 90\degree\ into the downstream section of the ion guide as illustrated in figure \ref{fig:RFQLength}.

\begin{figure}
    \centering
    \includegraphics[width=0.6\linewidth]{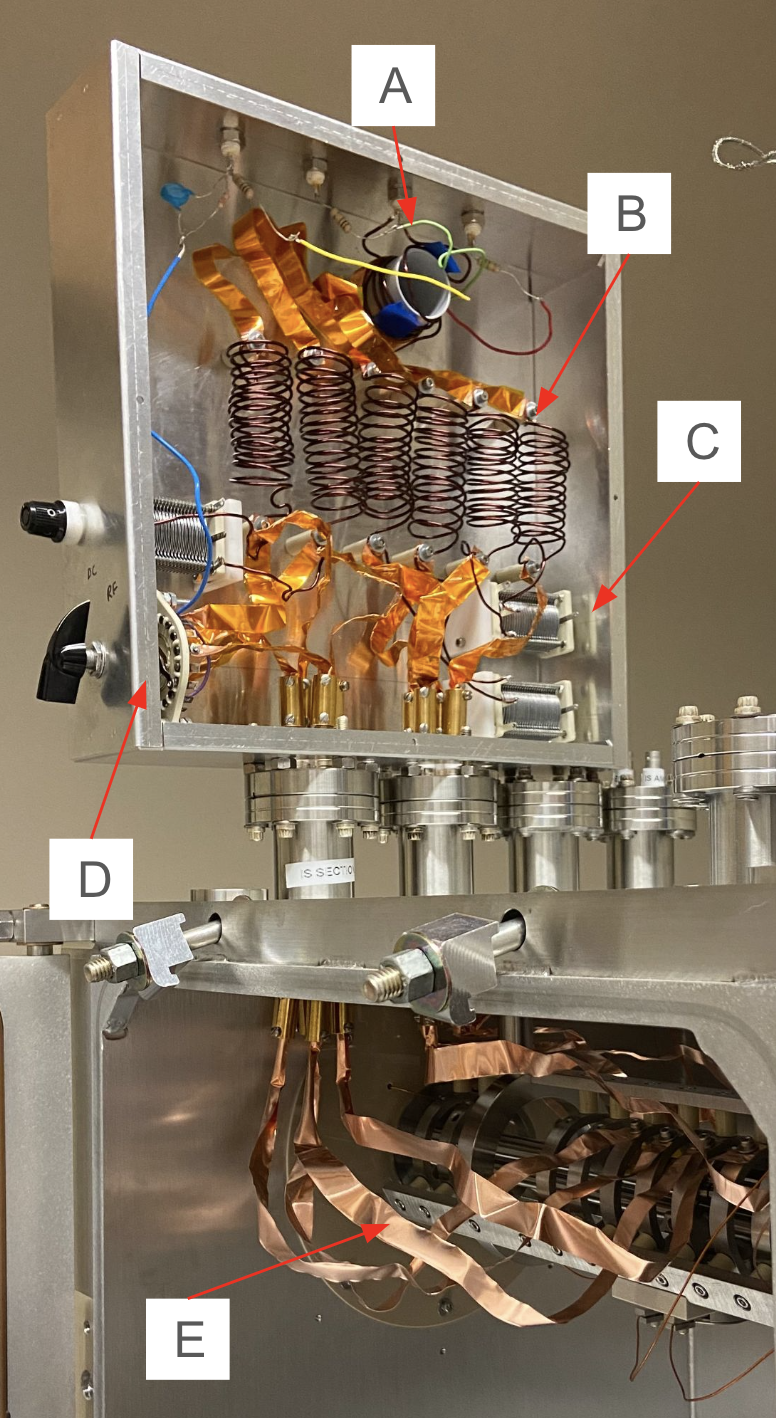}
    \caption{The ion guide RF circuit displaying A) an open air transformer leading into three separate circuits for each section of the ion guide. Each section contains B) two secondary inductors and C) a variable capacitor. D) A switch is used to add or remove RF from the lower two rods of the center section. E) Also shown is the ion guide mounted to the ceiling of its chamber.}
    \label{fig:circuit}
\end{figure}

Large amplitude RF signals are created by wrapping three secondary coils around a single primary inductor with a nominal transformer ratio of $8$, creating an open air transformer for precise $180\degree$ phase shifting between adjacent electrodes in each segment of the ion guide (see figure~\ref{fig:circuit}). Supplementary coils are used to increase the inductance of each section resulting in a resonant frequency for this system of 4.269 MHz. Each section includes a variable open-air capacitor, ranging from 12-300 pF, which allows for amplitude matching between subsequent sections. The DC potentials are coupled to the RF signal through the center tap on each of the three secondary coils of the transformer. This frequency, in combination with the amplitude of the RF signal covered, corresponds to a stability parameter $q$ \cite{WerthSpringer} of 0.1 - 0.3 for $^{39}$K, well within the region of stability for ion transport. Both the resonant frequency and power delivered to the circuit are stable over several hours of observation. A sample of RF amplitudes and mismatch across all sections of the ion guide is given in table \ref{tab:RFAmp}.

\begin{table}[]
    \centering
    \small
    \begin{tabular}{|l|c|r|}
     \hline
     \textbf{IG Rods} & \textbf{$0\degree$ Mode (V)} & \textbf{$90\degree$ Mode (V)}\\
     \hline
     \hline
     Upstream & $292(3.1\%)$ & $290(2.0\%)$ \\
     \hline
     Center Top & $280(0.5\%$ & $292(5.5\%)$ \\
     \hline
     Center Bottom & $290(6.5\%)$ & $0.9(3.3\%)$\\
     \hline
     Downstream & $284(6.5\%)$ & $284(6.5\%)$ \\
     \hline
     \end{tabular}
     
    \caption{Average measured amplitudes of the RF signal for each section of the ion guide at a generated RF power of 15 W. Values in parentheses are the normalized measurement range for each section.}
    \label{tab:RFAmp}
\end{table}

\section{Off-line Commissioning Set Up}
\label{set up}

\begin{table}[]
    \centering
    \small
    \begin{tabular}{|l|r|}
     \hline
     \textbf{Element} & \textbf{Setting}\\
     \hline
     \hline
     \multicolumn{2}{|l|}{\textbf{Constant Settings}}\\
     \hline
     Carpet Chamber Pressure & $2.25$ Torr\\
     \hline
     IG Chamber Pressure & $2.30\times10^{-3}$ Torr\\
     \hline
     Carpet Potential &  $0$ V\\
     \hline
     Carpet RF &  $12.45$ MHz\\
     \hline
     Carpet RF Amplitude &  $100$ V\\
     \hline
     Carpet LF & $50$ kHz\\
     \hline
     Carpet LF Amplitude & $5.0$ V\\
     \hline
     Filament Current & $2.10$ A\\
     \hline
     Potential Across Filament & $4.6$ V\\
     \hline
     Floating Potential of Filament & $130.1$  V\\
     \hline
     Anode Potential &  $80$ V\\
     \hline
   \multicolumn{2}{|l|}{\textbf{Initial Potentials}}\\
     \hline
     IG RF Amplitude & $286$ V\\
     \hline
     Carpet Aperture  & $-2$ V\\
     \hline
     Center $-$ Upstream & $0$ V\\
     \hline
     Downstream $-$ Center & $-100$ V\\
     \hline
     IG Aperture $-$ Downstream & $-100$ V\\
     \hline
     Lens $-$ IG Aperture &  $-20$ V\\
     \hline
     FC $-$ Lens & $-20$ V\\
     \hline
    \end{tabular}
    \caption{Settings for beam transport from the ion source at 0\degree\ that remained constant for every optimization scan in the off-line commissioning and initial settings used for the optimization scans. Element entries marked with a dash refer to potential differences between the two named electrodes.}
    \label{tab:0degsettings}
\end{table}

Off-line commissioning of the RFQ ion guide was performed for both the 0\degree\ and 90\degree\ configurations, which both utilize a similarly designed source holder for a thermionic emission potassium source. A description of the 0\degree\ source holder can be found in \cite{Davis2022-Flow}.

\begin{figure}
    \centering
    \includegraphics[width=0.8\linewidth]{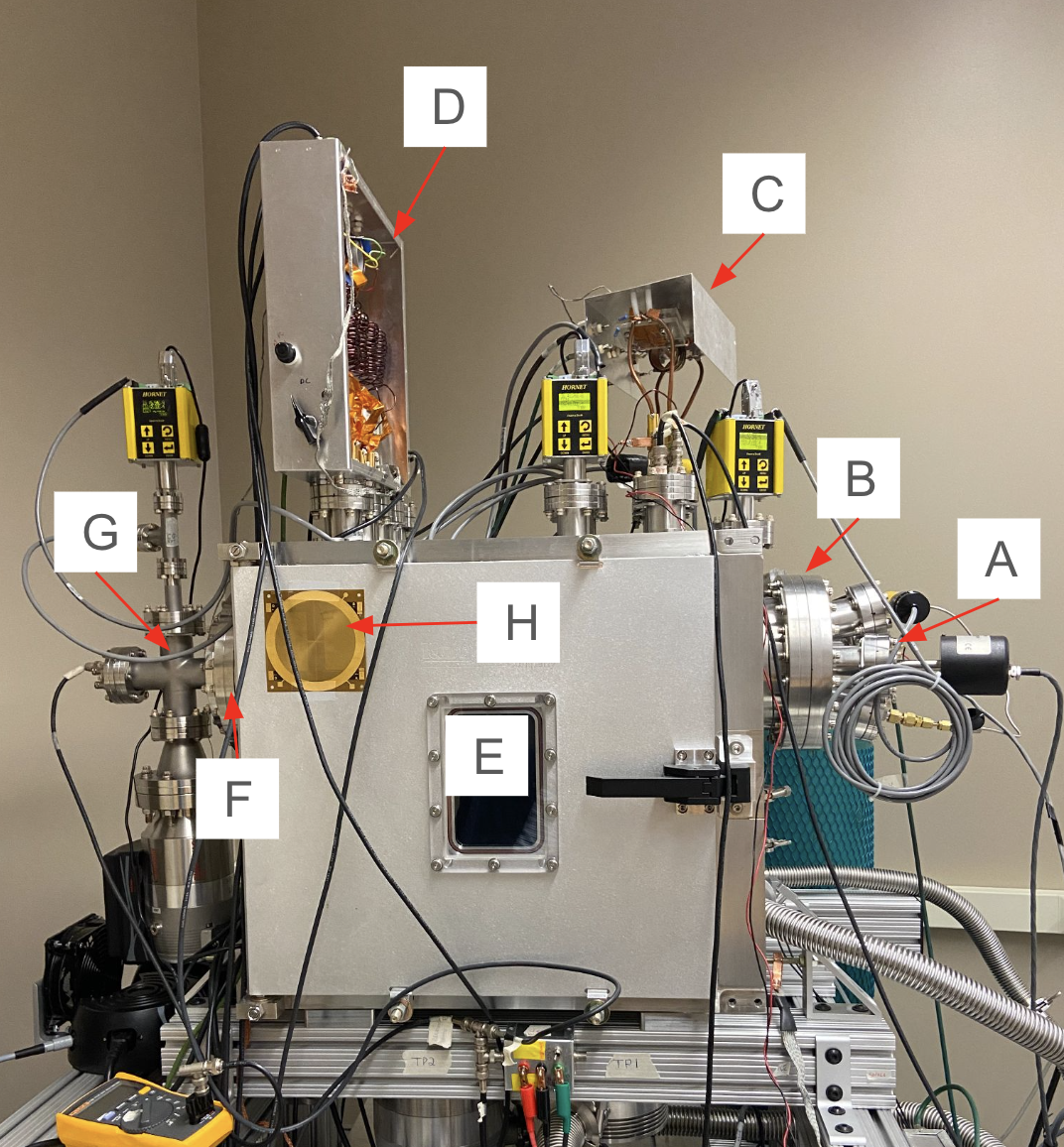}
    \caption{The off-line set up for the commissioning of the Ion Guide showing A) the flange containing the $^{39}$K source and the feed-through for the He gas. B) the nipple containing the RF carpet and C) the circuitry for the RF carpet. Also shown is D) the ion guide circuitry mounted in its final location and E) the location of the ion guide. Finally F) the nipple containing the two lens electrodes and G) the cross with a turbomolecular pump attached allowing for the evacuation of the chamber containing the Faraday cup. H) A copy of the RF carpet installed in the vacuum chamber.}
    \label{fig:setup}
\end{figure}

Off-line commissioning was completed in the final location of the ion guide, inside a custom made vacuum chamber (shown in figure~\ref{fig:setup}). This cube contains two ports located along the beam path. The upstream port (labeled B) contains an RF carpet mounted on a $1/16''$ thick PEEK disk with a $1/8''$ diameter hole at the center. This PEEK disk is mounted to the chamber wall, creating a pumping barrier between this smaller chamber and the adjacent volume. Attached to the back side of this PEEK disk is the RF carpet aperture, a $0.03''$ thick stainless steel (SS) plate with a $1/16''$ diameter aperture which protrudes into the hole of the PEEK disk to mitigate charging of the PEEK barrier when transporting ions. The distance between the RF carpet aperture and the upstream rods is around $3$ mm.

Mounted in location (E) of figure~\ref{fig:setup} is the ion guide itself. It is mounted on the ceiling of the chamber as shown in figure~\ref{fig:circuit}, such that the beam axis aligns with the central axis of the ports on either side. This chamber is kept at pressures on the order of $10^{-2}$ Torr using a 250 L/s and a 350 L/s turbo pump. Mounted to the downstream wall of the chamber is a second PEEK barrier that separates it from the downstream port. Attached to each side of this barrier are SS plates with a $1/16''$ diameter aperture. A small SS ring connects these two plates, providing a constant potential all along the ion guide (IG) aperture. In the downstream port (labeled F), there are two $2\,\sfrac{3}{16}''$ diameter drift tube electrodes spaced $1/4''$ apart, the first of which will be referred to as the lens. Finally, mounted on a flange located on the left most nipple of the cross labeled G, is a $1/16''$ thick stainless steel plate that sits directly against the most downstream electrode. Moving forward, the combination of this plate and the second drift tube electrode will be referred to as the Faraday cup (FC). This final chamber is pumped using a 150 L/s turbo pump to maintain pressures near $10^{-5}$ Torr.

The 0\degree\ configuration uses a source mounted on the flange labeled A in figure~\ref{fig:setup}. This configuration feeds in high purity He gas through a needle valve on the same flange. This gas is then evacuated through the central hole of the RF carpet into the ion guide chamber and pumped away. A pressure of $2.25$ Torr in the RF carpet chamber, corresponding to a pressure of $2.30\times10^{-3}$ Torr in the ion guide chamber, was maintained for most measurements. This configuration utilizes the RF carpet to transport ions from the source into the ion guide. Commissioning in the 90\degree\ configuration was done under vacuum and pressures in all chambers were on the order of $10^{-8}$ Torr. The source used in this configuration was mounted on the ion guide as described in section \ref{electrodeGeometry}. For all measurements in both configurations, the source current was maintained between $30-50$ pA. The uncertainty in current readings is dominated by fluctuations in the electrometer displayed current and typically ranges from $0.5-2$ pA, which is larger than the instrumental uncertainty of the electrometer.

Because the ionization potential of potassium ions ($4.34$ eV) is less than twice the work function of stainless steel ($4.08$ - $4.19$ eV range), the so-called potential electron emission from the Auger transition \cite{Hagstrum54} of the incoming ion near the surface can be neglected. However, low-energy ion-induced impact emission of electrons can still occur, and the yield depends on the cleanliness and oxygen level on the surface \cite{Walton99}. While our stainless-steel surface has been thoroughly cleaned to UHV standards, the level of oxygen on the surface has not been assessed. Hence, a conservative uncertainty of $2\%$, which corresponds to the electron production yield of a $250$ eV beam on an oxygen-contaminated surface has been added in quadrature to the FC current measurements.

\section{0\degree\ Source Results}
\label{0degResults}

In order to measure a transport efficiency, ions were sent to the upstream rods of the ion guide, via the RF carpet, with $-2$ V on the aperture and $-100$ V on the upstream rods. All electrodes after the upstream rods were grounded to ensure the ions are reflected back towards the upstream rods. After the current at this location was recorded, the bias on a given electrode was scanned and the current on the FC ($I_{FC}$) was recorded for each setting. The current at the upstream rods was then recorded again. Reported efficiencies are calculated as $\varepsilon = I_{FC}/I_{upstream}$, where $I_{upstream}$ is the linearly interpolated current on the upstream rods at the time the FC current was measured. The settings that did not vary between scans, as well as the initial potential difference between electrodes, can be found in table \ref{tab:0degsettings}. The RF carpet settings chosen were the ones that led to an efficient transport during its commissioning \cite{Davis2022-Flow}.

\begin{figure}
    \centering
    \includegraphics[width=\linewidth]{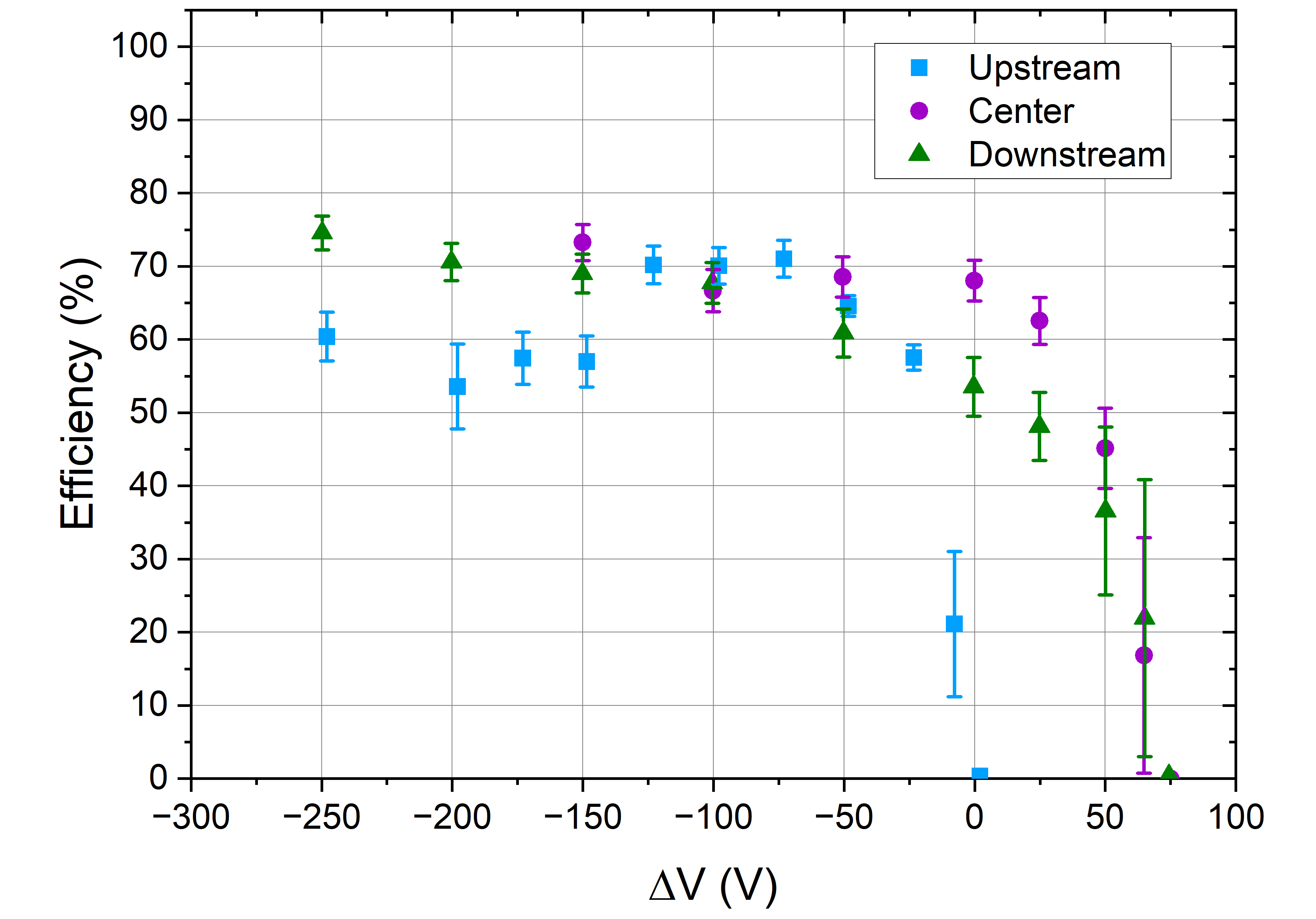}
    \caption{Transport efficiency of the beam produced from the ion source located at 0\degree\ as a function of $\Delta$V for each section of the ion guide. $\Delta$V corresponds to the potential difference between the named electrode and the electrode immediately upstream.}
    \label{fig:0degRods}
\end{figure}

Electrodes were optimized one at a time beginning with the upstream rods and proceeding downstream. Results from the optimization of the potential applied on each section of the ion guide are shown in figure~\ref{fig:0degRods}. It is clear the upstream section needs to be at a lower potential than the RF carpet aperture to efficiently transport the ions. However, for potential differences greater than $100$ V, the efficiency appears to slowly drop. It should be noted that this feature was reproducible. Hence, an optimal potential difference of $75$ V was chosen between this section and the RF carpet aperture. The center and downstream sections, on the other hand, need to be at a potential $75$ V higher than the previous section to block the beam. This corresponds to the beam energy set by the potential difference between the aperture and upstream section. As the potential on the center and downstream section gets more negative, the transport efficiency increases before plateauing. This plateau is reached at a potential difference around $0$ V and $-100$ V for the center and downstream sections respectively.

\begin{figure}
    \centering
    \includegraphics[width=\linewidth]{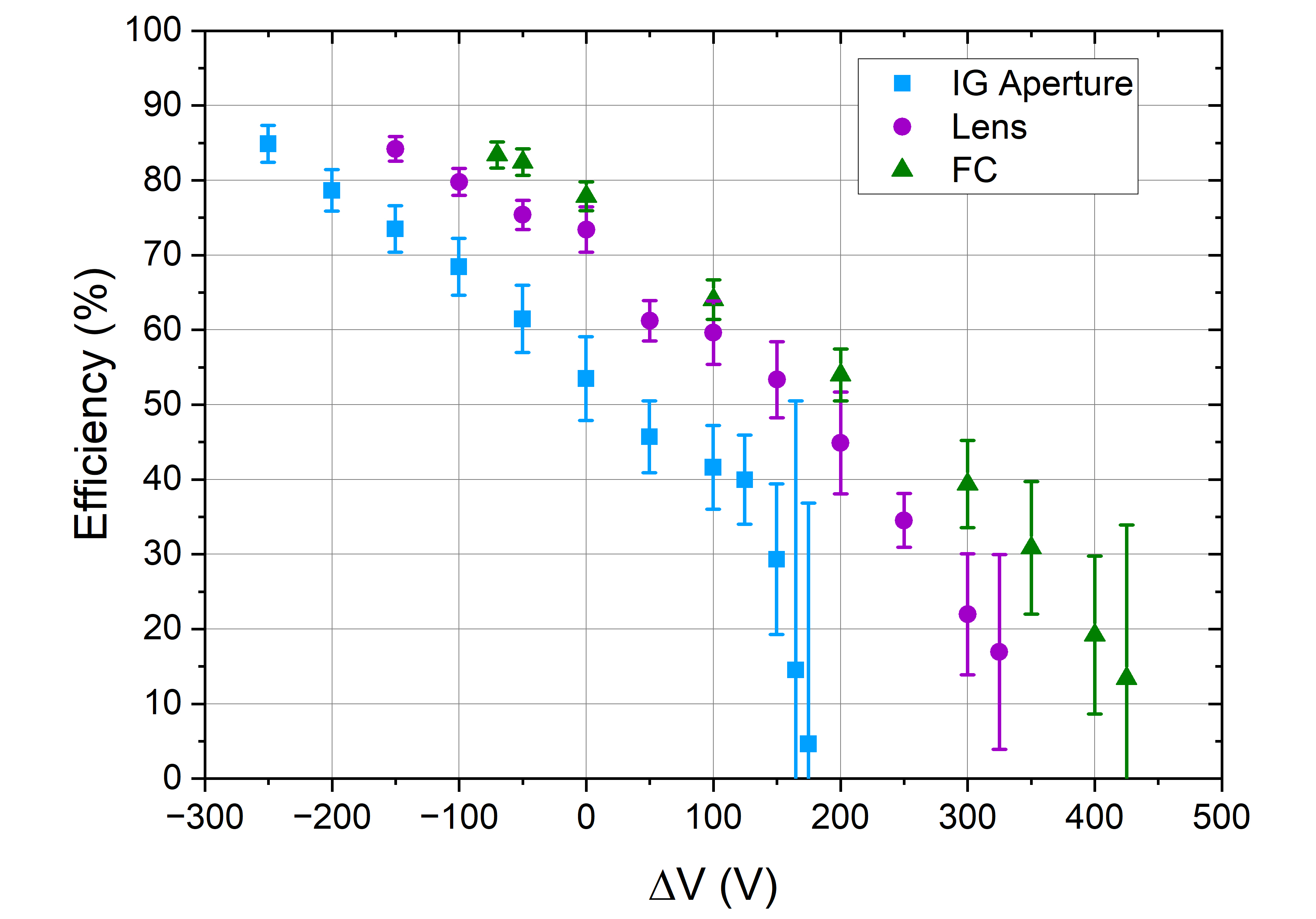}
    \caption{Transport efficiency of the beam produced from the ion source located at 0\degree\ as a function of $\Delta$V for the ion guide aperture, FC, and lens electrode. $\Delta$V corresponds to the difference in voltage between the named electrode and the electrode immediately upstream.}
    \label{fig:0degOptics}
\end{figure}

Optimization of the elements downstream of the ion guide are shown in figure~\ref{fig:0degOptics}. For each electrode downstream of the ion guide, a potential roughly corresponding to the beam energy is needed to block the ions. As the potential becomes more negative for each of these electrodes, transport efficiency increases. Current reading on biased electrodes was accomplished by also floating the inner shield of the triaxial connection going to the electrometer, which is rated up to a maximum of $500$ V, limiting the range of potential scanned. This was a limiting factor in optimizing the potential on electrodes downstream of the ion guide itself which is demonstrated by the lack of a plateau behavior in figure~\ref{fig:0degOptics}.

\begin{figure}
    \centering
    \includegraphics[width=\linewidth]{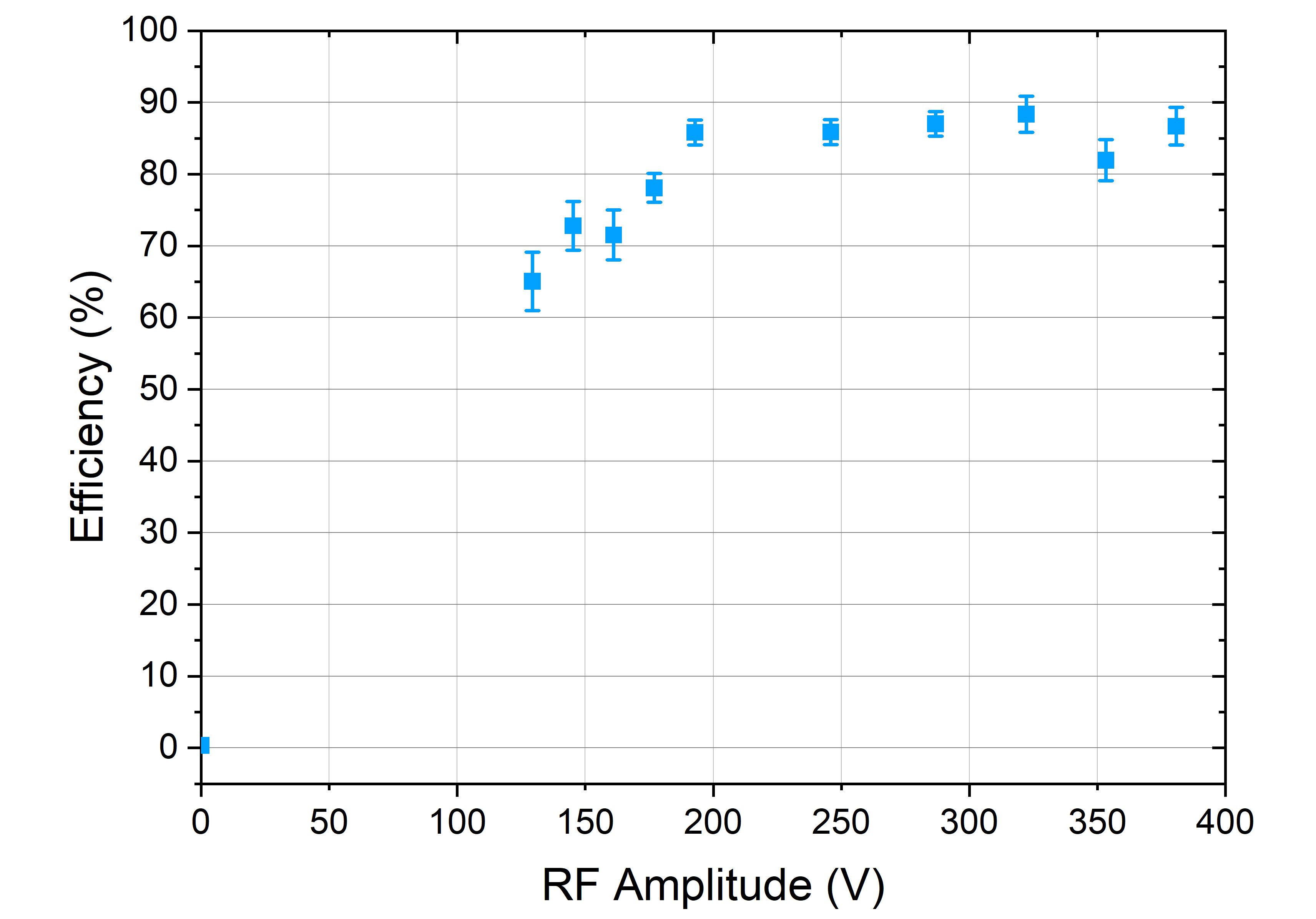}
    \caption{Transport efficiency of the beam produced from the ion source located at 0\degree\ as a function of the amplitude of the RF signal delivered to the rods of the ion guide. The first $6$ data points represent a $1$ W step in output power from the RF generator from $0$ W to $5$ W while the remaining points represent $5$ W steps from $5$ W to $30$ W.}
    \label{fig:0degRF}
\end{figure}

\begin{figure}
    \centering
    \includegraphics[width=\linewidth]{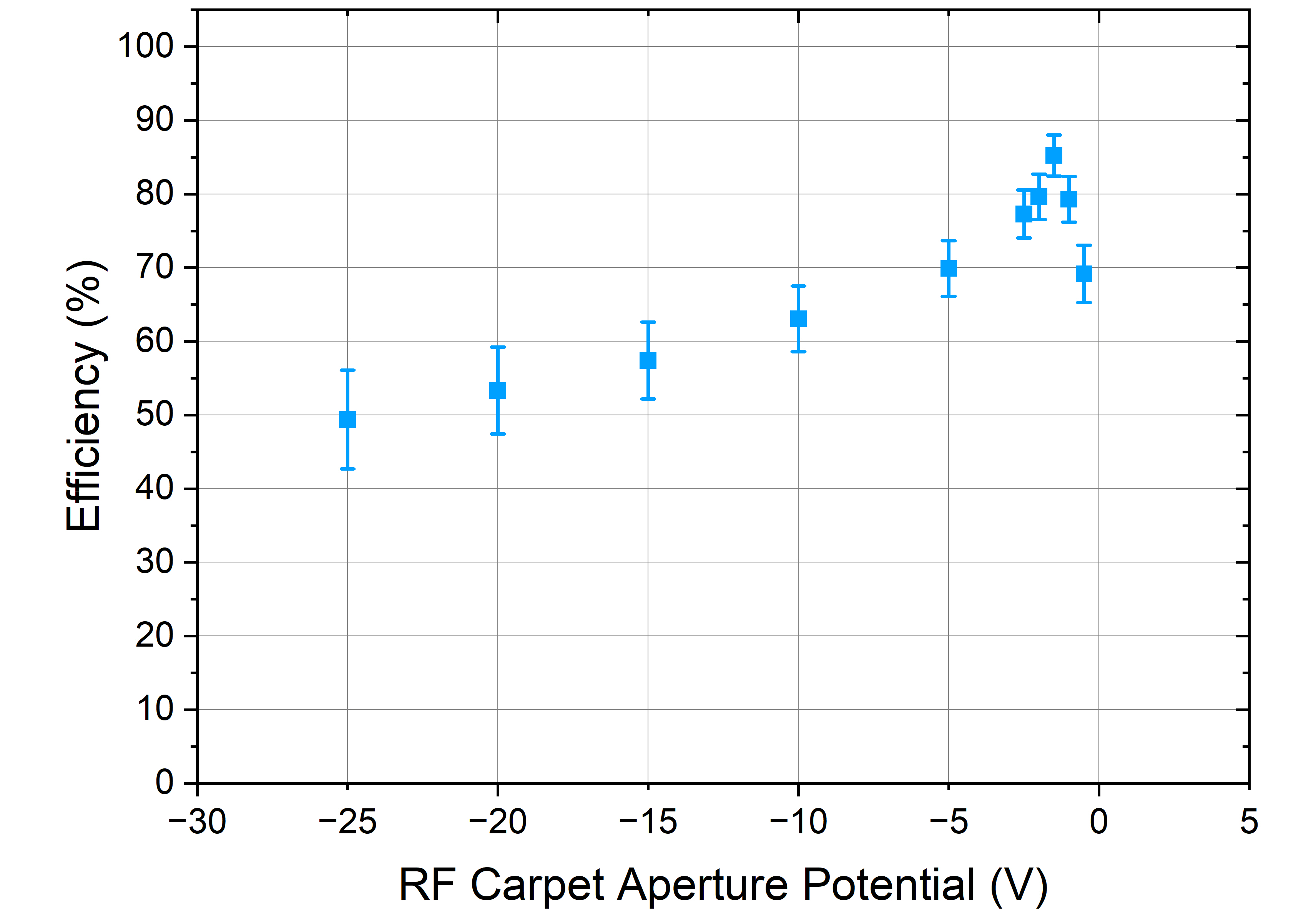}
    \caption{Transport efficiency of the beam produced from the ion source located at 0\degree\ as a function of the potential on the RF carpet aperture.}
    \label{fig:0degRFAp}
\end{figure}

After the potential on each electrode was optimized, the amplitude of the RF signal on the ion guide was scanned by changing the output power of the RF generator. Figure~\ref{fig:0degRF} shows that once an amplitude greater than $190$ V is reached, there is no improvement to the transport efficiency of ions through this region. Finally, the bias on the carpet aperture was optimized and the results are shown in figure~\ref{fig:0degRFAp}. Clearly the potential difference between the carpet aperture and the carpet needs to be quite small but nonzero. Once a potential difference greater than $1.5$ V is reached, transport efficiency begins to decrease. A summary of the full optimized settings can be found in table \ref{tab:0degOptimal}. 

\begin{table}[]
    \centering
    \small
    
    \begin{tabular}{|l|r|}
     \hline
     \textbf{Element} & \textbf{Optimized Setting}\\
     \hline
     \hline
     Carpet Aperture  & $-1.5$ V\\
     \hline
     Upstream Rods & $-75$ V\\
     \hline
     Center Rods & $-75$ V\\
     \hline
     Downstream Rods & $-175$ V\\
     \hline
     IG Aperture & $-275$ V\\
     \hline
     Lens &  $-475$ V\\
     \hline
     FC & $-495$ V\\
     \hline
     RF Amplitude & $200$ V\\
     \hline
    \end{tabular}
    \caption{0\degree\ source settings for optimal transport of ions through the ion guide.}
    \label{tab:0degOptimal}
\end{table}

\subsection{Effect of Pressure}

\begin{figure}
    \centering
    \includegraphics[width=\linewidth]{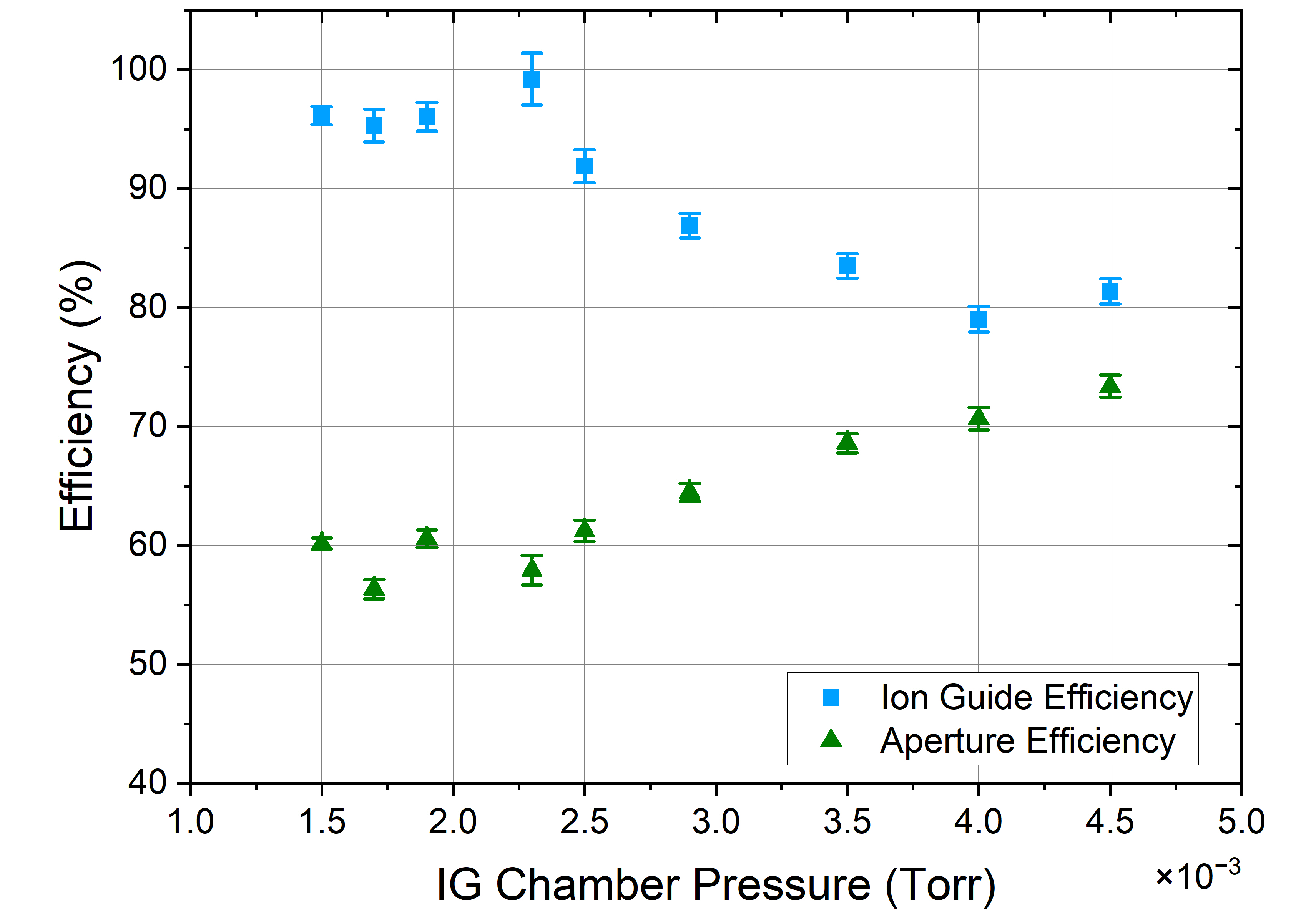}
    \caption{Transport efficiency of the beam produced from the ion source located at 0\degree\ as a function of pressure in the IG chamber. The blue squares represent the transport efficiency of the ion guide alone while the green triangles show the transport efficiency to the aperture between the RF carpet and IG chambers. It should be noted that pressures of $1.5\times10^{-3}$ and $4.5\times10^{-3}$ Torr correspond to pressures of 1.3 and 5.3 Torr respectively in the RF carpet chamber.}
    \label{fig:0degPressure}
\end{figure}

The effect of pressure in the ion guide chamber on the transport of ions through the ion guide and acceptance through the carpet aperture and into the ion guide was studied by varying the He pressure in the carpet chamber. At each pressure, the ions were first transported to the outmost electrode of the RF carpet (the ring electrode) and the current on the ring was recorded after which the current on the upstream rods and FC was recorded. Before moving onto another pressure, the transported beam current to the ring electrode was measured again. A linear interpolation of the current measured on the ring electrode was used to calculate the transport efficiency to the carpet aperture at the time the upstream current was recorded, $\varepsilon_{aperture} = I_{upstream}/I_{ring}$. To calculate the ion guide efficiency as $\varepsilon_{IG} = I_{FC}/I_{upstream}$, $\varepsilon_{aperture}$ was used to determine the upstream current at the time the FC current was measured. Results from these scans (see figure~\ref{fig:0degPressure}) demonstrate that while higher pressures correlate to better transport through the carpet aperture--showing a roughly $10$\% increase in transport efficiency across this pressure range--lower pressures are preferred to optimize transport through the ion guide itself. More specifically, transport efficiency through the ion guide above 95\% was observed when the pressure is below $2.5\times10^{-3}$ Torr in the ion guide chamber. Simulation work is ongoing, aimed at improving the transport through the carpet aperture.

\section{90\degree\ Source Results}


\begin{table}[]
    \centering
    \small
    
    \begin{tabular}{|l|r|}
     \hline
     \textbf{Element} & \textbf{Setting}\\
     \hline
     \hline
     \multicolumn{2}{|l|}{\textbf{Constant Settings}}\\
     \hline
     IG Chamber Pressure & $6.13\times10^{-8}$ Torr\\
     \hline
     Filament Current & $1.30$ A\\
     \hline
     Potential Across Filament& $2.7$ V\\
     \hline
     Floating Potential of Filament & $6.18$  V\\
     \hline
     Anode Potential &  $5.22$ V\\
     \hline
     Center Rods Potential & $0$ V\\
     \hline
     \multicolumn{2}{|l|}{\textbf{Initial Potentials}}\\
     \hline
     IG RF Amplitude & $288$ V\\
     \hline
     Upstream Rods & $30$ V\\
     \hline
     Downstream Rods & $-5$ V\\
     \hline
     IG Aperture $-$ Downstream Rods & $-195$ V\\
     \hline
     Lens $-$ IG Aperture & $-100$ V\\
     \hline
     Lens $-$ FC & $-150$  V\\
     \hline
    \end{tabular}
    \caption{Settings for beam transport from the 90\degree\ source that remained constant for every optimization scan in the off-line commissioning as well as initial biases used for subsequent electrodes. Element entries marked with a dash refer to potential differences between the two named electrodes.}
    \label{tab:90degsettings}
\end{table}

\begin{figure}
    \centering
    \includegraphics[width=\linewidth]{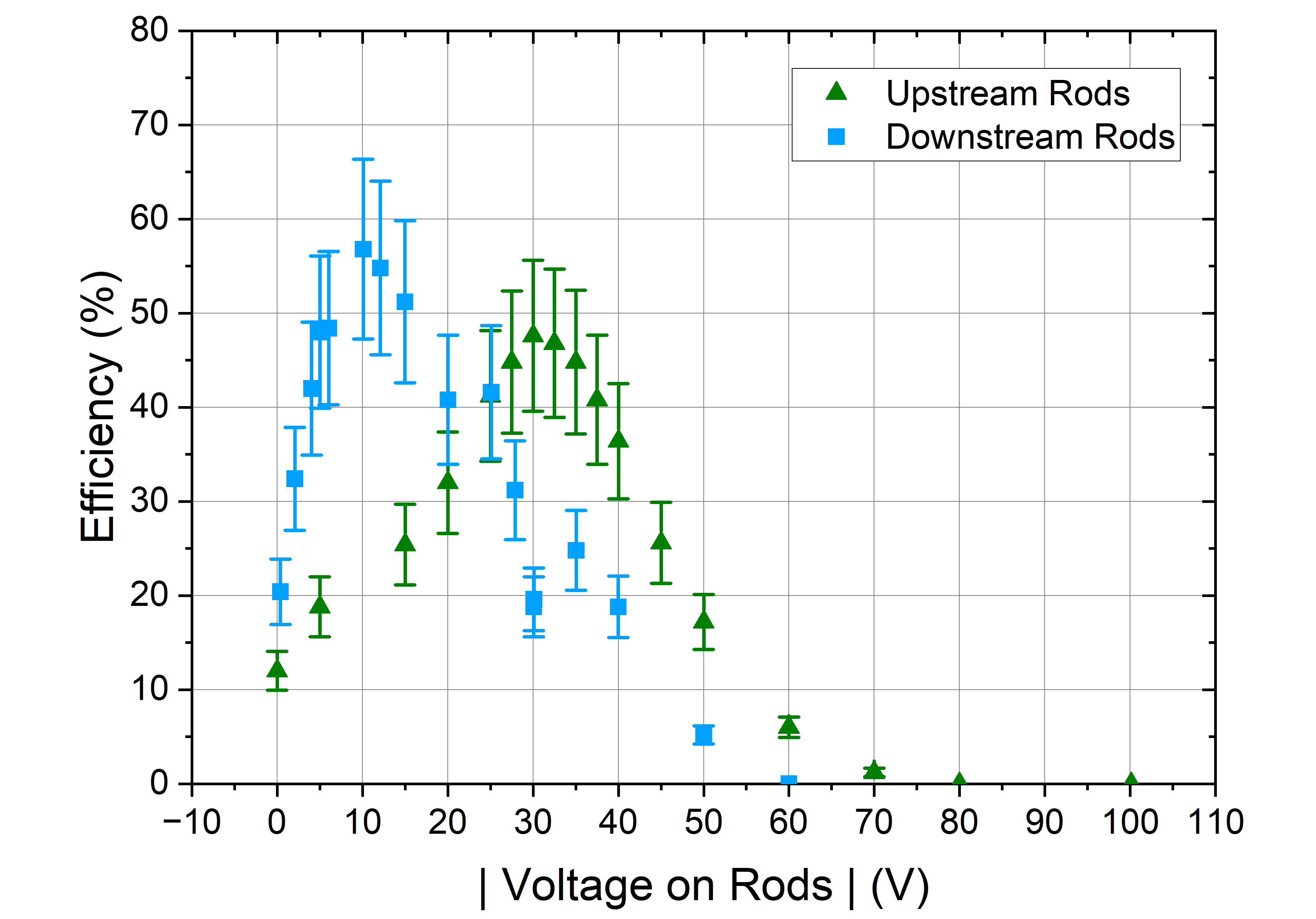}
    \caption{Transport efficiency of the beam produced from the ion source located at 90\degree\ as a function of the absolute value of the potential on upstream rods (in green) and downstream rods (in blue) of the ion guide. The potential applied to the upstream rods was positive, while the downstream rods were biased negatively. The central section was at ground potential.}
    \label{fig:90degRods}
\end{figure}

The transport efficiency of the ion guide using ions from the 90\degree\ source was ascertained by first recording the current on all the rods of the ion guide together ($I_{rods}$). This was done with all rods of the ion guide set to $0$ V and the source and anode at $6$ and $5$ V respectively. Capacitive couplings between several rods within the ion guide caused large variations in the current reading and this is reflected in the much larger uncertainties reported for the 90\degree\ configuration. Once the current on the rods was recorded, each element was scanned and the current on the FC was recorded. Efficiencies for this configuration are reported as $\varepsilon = I_{FC}/I_{rods}$. The center rods were maintained at $0$V for all optimization scans. Optimization was finalized by scanning the RF forward power on the ion guide. Finally, the current on the rods together was confirmed. Initial settings and elements that were not varied are recorded in table \ref{tab:90degsettings}. 

Results for the optimization of the upstream and downstream rods are shown in figure~\ref{fig:90degRods}. While maintaining a potential of $-5$ V on the downstream rods, the potential on the upstream rods was scanned. At $0$ V on the upstream rods, most of the ions do not make it to the FC. As the potential becomes more positive, transport efficiency increases to a clear optimal potential of $30$ V. Beyond this, transport efficiency decreases. Leaving the upstream rods at this new optimized potential while scanning the downstream rods provided a similar trend. Where $0$ V corresponds to the stopping of most ions and a clear optimal potential is observed at $-10$ V. As the potential on the downstream rods becomes more negative, ion transport decreases.

\begin{figure}
    \centering
    \includegraphics[width=\linewidth]{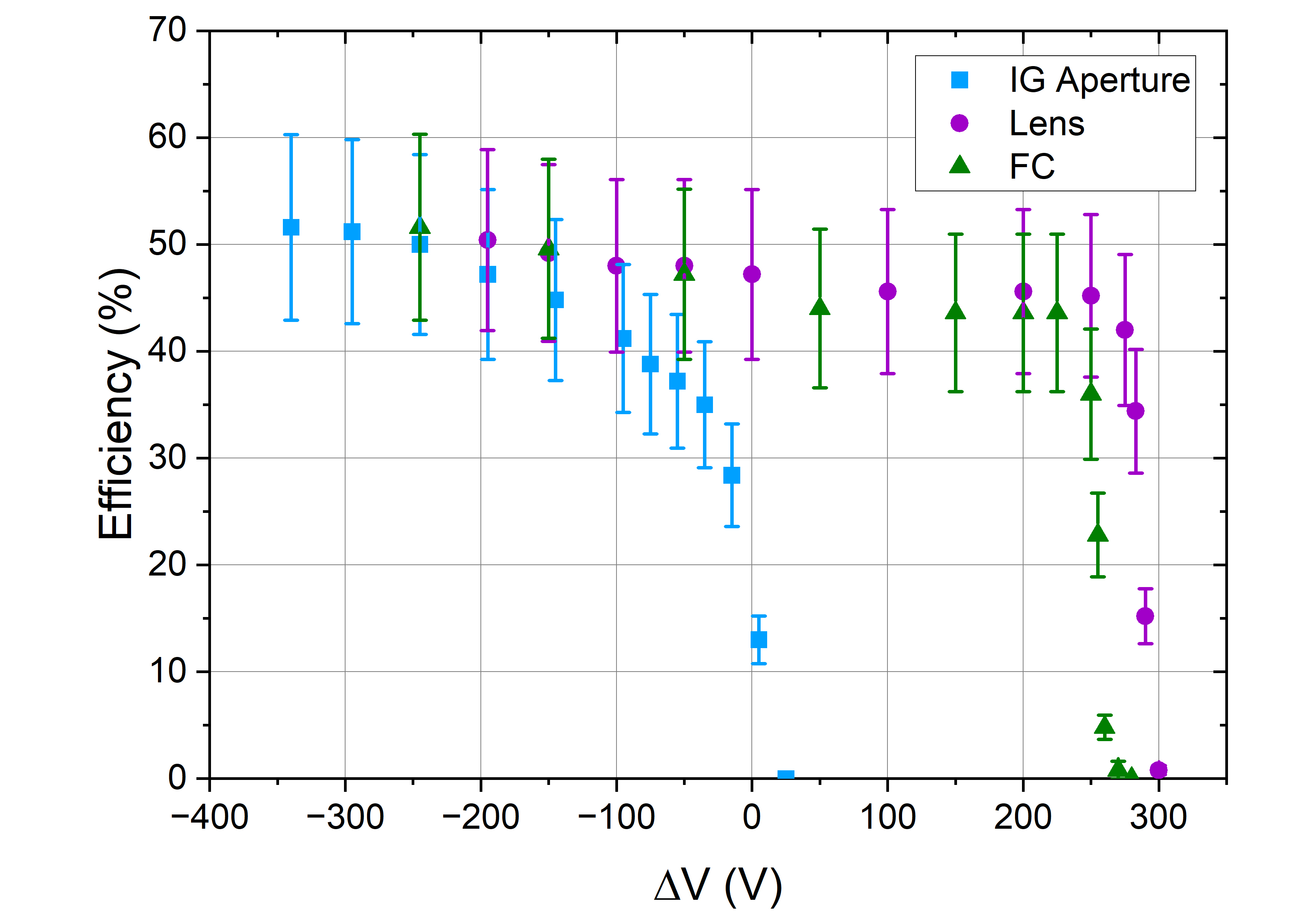}
    \caption{Transport efficiency of the beam produced from the ion source located at 90\degree\ as a function of $\Delta$V for the ion guide aperture, lens electrode and FC. $\Delta$V corresponds to the difference in voltage between the named electrode and the electrode immediately upstream.}
    \label{fig:90degOptics}
\end{figure}

Once the potential of the rods of the ion guide had been optimized, the electrodes downstream were scanned and these results can be seen in figure~\ref{fig:90degOptics}. A potential, which roughly corresponds to the beam energy, of roughly $25$ V on the IG aperture is needed to stop the beam. As the potential on the aperture becomes more negative, transport efficiency increases before plateauing for potential differences greater than $-250$ V. In contrast, the lens and FC electrodes do not need a lower potential than the preceding electrode to transport ions. Specifically, these electrodes can transport ions at an efficiency over $45$\% with a potential $250$ V higher than the electrode immediately upstream. However, transport efficiency slowly increases once the potential difference becomes negative. Due to the same constraints discussed in section \ref{0degResults}, the potentials needed to optimize these electrodes could not be reached.

Finally, the amplitude of the RF wave was scanned. Amplitudes $125$ V and lower are not sufficient enough to confine ions coming from this 90\degree\ turn. The transport efficiency climbs as the amplitude is increased until reaching a plateau at $280$ V. A summary of the final optimal settings for this configuration can be found in table \ref{tab:90degOptimal}. Total efficiency of ions that make it to the Faraday cup from the 90\degree\ source is $60(10)$\%. 

\begin{table}[]
    \centering
    \small
    
    \begin{tabular}{|l|r|}
     \hline
     \textbf{Element} & \textbf{Optimized Setting}\\
     \hline
     \hline
     Upstream Rods & $30$ V\\
     \hline
     Center Rods & $0$ V\\
     \hline
     Downstream Rods & $-10$ V\\
     \hline
     IG Aperture & $-255$ V\\
     \hline
     Lens & $-255$ V\\
     \hline
     FC & $-305$  V\\
     \hline
     IG RF Amplitude & $200$ V\\
     \hline
    \end{tabular}
    \caption{90\degree\ source settings for optimal transport from the 90\degree\ source.}
    \label{tab:90degOptimal}
\end{table}

\begin{figure}
    \centering
    \includegraphics[width=\linewidth]{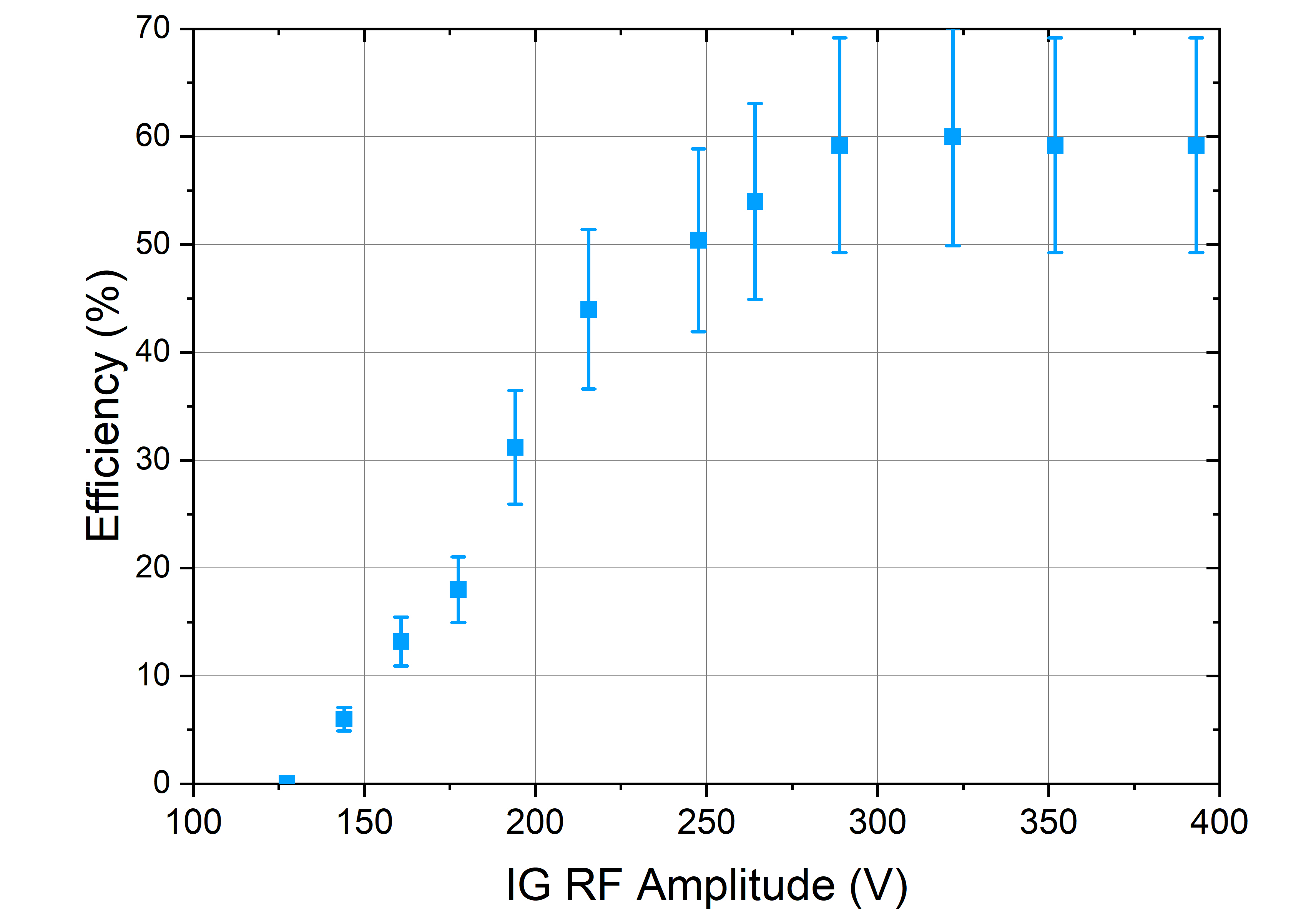}
    \caption{Transport efficiency of the beam produced from the ion source located at 90\degree\ as a function of the forward power on the RF generator for the ion guide.}
    \label{fig:90degRFFP}
\end{figure}

\section{Conclusion and Outlook}

The St.~Benedict ion guide has been fully commissioned using an off-line source. In the 0\degree\ configuration, over $95$\% of ions that make it into the ion guide are transported through. Simulations are ongoing to understand the observed loss of ions in the carpet aperture that results in an overall transport efficiency of $60$\%. In the 90\degree\ configuration, over 60\% of ions make it through the ion guide which is sufficient for the purposes of this source. Implementation of a 90\degree\ source in such a way as is outlined in this paper is a valid option for testing and commissioning of downstream elements of St.~Benedict in the FWTR. 

St.~Benedict is currently coming online in order to extract $\rho$ for nuclear mirrors including $^{11}$C, $^{13}$N, $^{15}$O, $^{17}$F, $^{19}$Ne, $^{21}$Na, $^{25}$Al, and $^{41}$Sc. These measurements will both greatly expand the list of nuclei for which $\rho$ is known and ultimately shed light on the unitarity of the CKM matrix.

\section*{Acknowledgment}

The authors would like to thank M.~Sanford for the designing and machining of the ion guide.
This work was conducted with the support of the National Science Foundation under Grant PHY-1725711, PHY-2011890, and of the University of Notre Dame.


\begin{thebibliography}{99}
   
\bibitem{PDG}
M. ~Tanabashi, et al., \href{https://doi.org/10.1103/PhysRevD.98.030001}{Review of particle physics}, Phys. Rev. D 98 (2018) 030001.
\newblock \href{https://doi.org/10.1103/PhysRevD.98.030001}{doi: 10.1103/PhysRevD.98.030001}.
\newline\urlprefix\url{https://link.aps.org/doi/10.1103/PhysRevD.98.030001}

\bibitem{PhysRevD.86.010001}
J. ~Beringer, et al., \href{https://doi.org/10.1103/PhysRevD.86.010001}{Review of particle physics},  Phys. Rev. D 86 (2012) 010001.
\newblock \href{https://doi.org/10.1103/PhysRevD.86.010001}{doi: 10.1103/PhysRevD.86.010001}.
\newline\urlprefix\url{https://link.aps.org/doi/10.1103/PhysRevD.86.010001}

\bibitem{Severijns_2013}
N.~Severijns, O.~Navilliat-Cuncic, \href{https://doi.org/10.1088/0031-8949/2013/T152/014018}{Structure and symmetries of the weak interaction in nuclear beta decay}, Physics Scripta T152 (2013) 014018.
\newblock \href{https://doi.org/10.1088/0031-8949/2013/T152/014018}{doi: 10.1088/0031-8949/2013/T152/014018}.
\newline\urlprefix\url{https://dx.doi.org/10.1088/0031-8949/2013/T152/014018}

\bibitem{3sigma}
V.~Cirigliano and A.~Crivellin and M.~Hoferichter,  M.~Moulson, \href{https://doi.org/10.1016/j.physletb.2023.137748}{Scrutinizing CKM unitarity with a new measurement of the $K_{mu3}/K_{mu2}$ branching fraction}, Physics Letters B 838 (2023) 137748.
\newblock \href{https://doi.org/10.1016/j.physletb.2023.137748}{doi: 10.1016/j.physletb.2023.137748}.
\newline\urlprefix\url{https://www.sciencedirect.com/science/article/pii/S0370269323000825}

\bibitem{Brodeur2016-StBenedict}
M.~Brodeur, J.~Kelly, J.~Long, C.~Nicoloff, B.~Schultz,
  \href{http://www.sciencedirect.com/science/article/pii/S0168583X15012847}{Determination
  from light nuclide mirror transitions}, Nuclear Instruments and Methods in
  Physics Research Section B: Beam Interactions with Materials and Atoms 376
  (2016) 281 -- 283, proceedings of the \{XVIIth\} International Conference on
  Electromagnetic Isotope Separators and Related Topics (EMIS2015), Grand
  Rapids, MI, U.S.A., 11-15 May 2015.
\newblock \href{https://doi.org/10.1016/j.nimb.2015.12.038}{doi:10.1016/j.nimb.2015.12.038}.
\newline\urlprefix\url{http://www.sciencedirect.com/science/article/pii/S0168583X15012847}

\bibitem{OMalley2020-StBenedict}
P.~O'Malley, M.~Brodeur, D.~Burdette, J.~Klimes, A.~Valverde, J.~Clark,
  G.~Savard, R.~Ringle, V.~Varentsov,
  \href{https://www.sciencedirect.com/science/article/pii/S0168583X19302010}{Testing
  the weak interaction using St.~Benedict at the University of Notre Dame},
  Nuclear Instruments and Methods in Physics Research Section B: Beam
  Interactions with Materials and Atoms 463 (2020) 488--490.
\newblock \href{https://doi.org/10.1016/j.nimb.2019.04.017}{doi: 10.1016/j.nimb.2019.04.017}.
\newline\urlprefix\url{https://www.sciencedirect.com/science/article/pii/S0168583X19302010}

\bibitem{Porter2023-StBenedict}
W.~S. Porter, D.~W. Bardayan, M.~Brodeur, D.~P. Burdette, J.~A. Clark, A.~T.
  Gallant, A.~M. Houff, J.~J. Kolata, B.~Liu, P.~D. O’Malley, C.~Quick,
  F.~Rivero, G.~Savard, A.~A. Valverde, R.~Zite,
  \href{https://www.mdpi.com/2218-2004/11/10/129}{The st. benedict facility:
  Probing fundamental symmetries through mixed mirror $\beta$-decays}, Atoms
  11~(10) (2023).
\newblock \href{https://doi.org/10.3390/atoms11100129}{doi:10.3390/atoms11100129}.
\newline\urlprefix\url{https://www.mdpi.com/2218-2004/11/10/129}

\bibitem{SAVARD2020258}
G.~Savard, M.~Brodeur, J.~A. Clark, R.~A. Knaack, A.~A. Valverde, 
  \href{https://www.sciencedirect.com/science/article/pii/S0168583X19303295}{The N~=~126 factory: A new facility to produce very heavy neutron-rich isotopes}, Nuclear Instruments and Methods in Physics Research Section B: Beam Interactions with Materials and Atoms 463 (2020) 258--261.
\newblock \href{https://doi.org/10.1016/j.nimb.2019.05.024}{doi: 10.1016/j.nimb.2019.05.024}.
\newline\urlprefix\url{https://www.sciencedirect.com/science/article/pii/S0168583X19303295}

\bibitem{BLOCK20084521}
M.~Block, C.~Bachelet, G.~Bollen, M.~Facina, C.~M. Folden, C.~Gu\'enaut, A.~A.Kwiatkowski, D.~J. Morrissey, G.~K. Pang, A.~Prinke, R.~Ringle, J.~Savory, P.~Schury, S.~Schwarz, \href{https://www.sciencedirect.com/science/article/pii/S0168583X08007684}{Mass measurements of rare isotopes with the LEBIT facility at the NSCL} Nuclear Instruments and Methods in Physics Research Section B: Beam Interactions with Materials and Atoms 266 (19) (2008) 4521--4526, proceedings of the XVth International Conference on Electromagnetic Isotope Separators and Techniques Related to their Applications. 
\newblock \href{https://doi.org/10.1016/j.nimb.2008.05.098}{doi: 10.1016/j.nimb.2008.05.098}.
\newline\urlprefix\url{https://www.sciencedirect.com/science/article/pii/S0168583X08007684} 

\bibitem{BECCHETTI2003377}
F.~D. Becchetti, M.~Y. Lee, T.~W. O'Donnell, D.~A. Roberts, J.~J. Kolata, L.~O. Lamm,
  G.~Rogachev, V.~Guimar$\tilde{a}$es, P.~A. DeYoung, S.~Vincent,
  \href{http://www.sciencedirect.com/science/article/pii/S016890020301101X}{The
  twinsol low-energy radioactive nuclear beam apparatus: status and recent
  results}, Nuclear Instruments and Methods in Physics Research Section A:
  Accelerators, Spectrometers, Detectors and Associated Equipment 505 (2003)
  377 -- 380, proceedings of the tenth Symposium on Radiation Measurements and
  Applications.
\newblock \href{https://doi.org/10.1016/S0168-9002(03)01101-X}{doi:10.1016/S0168-9002(03)01101-X}.
\newline\urlprefix\url{http://www.sciencedirect.com/science/article/pii/S016890020301101X}

\bibitem{OMALLEY2016417}
P.~D. O'Malley, D.~W. Bardayan, J.~J. Kolata, M.~R. Hall, O.~Hall, J.~Allen, F.~D. Becchetti, \href{https://www.sciencedirect.com/science/article/pii/S0168583X15012793}{Upgrades for TwinSol facility}, Nuclear Instruments and Methods in Physics Research Section B: Beam Interactions with Materials and Atoms 376 (2016) 417--419, proceedings of the XVIIth International Conference on Electromagnetic Isotope Separators and Related Topics (EMIS2015), Grand Rapids, MI, U.S.A., 11-15 May 2015. 
\newblock \href{https://doi.org/10.1016/j.nimb.2015.12.033}{doi: 10.1016/j.nimb.2015.12.033}.
\newline\urlprefix\url{https://www.sciencedirect.com/science/article/pii/S0168583X15012793}

\bibitem{BRODEUR202379}
M.~Brodeur, T.~Ahn, D.~Bardayan, D.~Burdette, J.~Clark, A.~Gallant, J.~Kolata, B.~Liu, P.~O’Malley, W.~Porter, R.~Ringle, F.~Rivero, G.~Savard, A.~Valverde, R.~Zite,
\href{https://www.sciencedirect.com/science/article/pii/S0168583X23002215}{Construction of St.~Benedict}, Nuclear Instruments and Methods in Physics Research Section B: Beam Interactions with Materials and Atoms 541 (2023) 79--81.
\newblock \href{https://doi.org/10.1016/j.nimb.2023.05.024}{doi: 10.1016/j.nimb.2023.05.024}.
\newline\urlprefix\url{https://www.sciencedirect.com/science/article/pii/S0168583X23002215}

\bibitem{WerthSpringer}
G.~Werth, V.~N.~Gheorghe, F.~G.~Major,
\href{https://link.springer.com/book/10.1007/978-3-540-92261-2}{Charged particle traps II}, Springer (2009).
\newblock \href{https://doi.org/10.1007/978-3-540-92261-2}{doi: 10.1007/978-3-540-92261-2}.
\newline\urlprefix\url{https://link.springer.com/book/10.1007/978-3-540-92261-2}

\bibitem{Davis2022-Flow}
C.~Davis, R.~Bualuan, O.~Bruce, D.~Burdette, A.~Cannon, T.~Florenzo, D.~Gan, J.~Harkin, B.~Liu, J.~Long, P.~O’Malley, W.~Porter, F.~Rivero, M.~Yeck, R.~Zite, M.~Brodeur,
\href{https://www.sciencedirect.com/science/article/pii/S0168900222007148}{Commissioning of the St.~Benedict RF carpet}, Nuclear Instruments and Methods in Physics Research Section A: Accelerators, Spectrometers, Detectors and Associated Equipment 1042 (2022) 167422.
\newblock \href{https://doi.org/10.1016/j.nima.2022.167422}{doi: 10.1016/j.nima.2022.167422}.
\newline\urlprefix\url{https://www.sciencedirect.com/science/article/pii/S0168900222007148}

\bibitem{Davis2022-Static}
C.~Davis, O.~Bruce, D.~Burdette, T.~Florenzo, B.~Liu, J.~Long, P.~O’Malley,
  M.~Yeck, M.~Brodeur,
  \href{https://www.sciencedirect.com/science/article/pii/S0168900222001279}{Transport
  tests of the St.~Benedict first-stage extraction system}, Nuclear Instruments
  and Methods in Physics Research Section A: Accelerators, Spectrometers,
  Detectors and Associated Equipment 1031 (2022) 166509.
\newblock \href{https://doi.org/10.1016/j.nima.2022.166509}{doi: 10.1016/j.nima.2022.166509}.
\newline\urlprefix\url{https://www.sciencedirect.com/science/article/pii/S0168900222001279}

\bibitem{CoolerBuncher}
D.~P. Burdette, R.~Zite, M.~Brodeur, A.~A. Valverde, O.~Bruce, R.~Bauluan, A.~Cannon, J.~A. Clark, C.~Davis, T.~Florenzo, A.~T. Gallant, J.~Harkin, A.~M. Houff, J.~Li, B.~Liu, J.~Long, P.~D. O'Malley, W.~S. Porter, C.~Quick, R.~Ringle, F.~Rivero, G.~Savard, M.~A. Yeck,  \href{https://doi.org/10.1016/j.nima.2025.171187}{Off-line commissioning of the St. Benedict radiofrequency quadrupole cooler-buncher}, Nuclear Instruments and Methods in Physics Research Section A: Accelerators, Spectrometers, Detectors and Associated Equipment 1084 (2026) 171187. 
\newblock \href{https://doi.org/10.1016/j.nima.2025.171187}{doi: 10.1016/j.nima.2025.171187}.
\newline\urlprefix\url{https://www.sciencedirect.com/science/article/pii/S0168900225009891}

\bibitem{Douglas2002}
D.~J. Douglas, N.~V. Konenkov, \href{https://doi.org/10.1002/rcm.735}{Influence of the 6th and 10th spatial harmonics on the peak shape of a quadrupole mass filter with round rods}, Rapid Communications in Mass Spectrometry 16 (15) (2002) 1425--1431. 
\newblock \href{https://doi.org/10.1002/rcm.735}{doi: 10.1002/rcm.735}.
\newline\urlprefix\url{https://doi.org/10.1002/rcm.735}

\bibitem{Hagstrum54}
H.~D.~Hagstrum, \href{https://journals.aps.org/pr/abstract/10.1103/PhysRev.96.336}{Theory of auger ejection of electrons from metals by ions}, Phys. Rev. 96 (2) (1954) 336--365. 
\newblock \href{https://doi.org/10.1103/PhysRev.96.336}{doi: 10.1103/PhysRev.96.336}.
\newline\urlprefix\url{https://journals.aps.org/pr/abstract/10.1103/PhysRev.96.336}

\bibitem{Walton99}
S.~G.~Walton, J.~C.~Tucek, R.~L.~Champion, Y.~Wang, \href{https://pubs.aip.org/aip/jap/article-pdf/85/3/1832/19112809/1832_1_online.pdf}{Low energy, ion-induced electron and ion emission from stainless steel: The effect of oxygen coverage and the implications for discharge modeling}, Journal of Applied Physics 85 (3) (1999) 1832--1837
\newblock \href{https://doi.org/10.1063/1.369330}{doi: 10.1063/1.369330}
\newline\urlprefix\url{https://pubs.aip.org/aip/jap/article-pdf/85/3/1832/19112809/1832_1_online.pdf}
 
\end{thebibliography}
\end{document}